\documentclass[12pt]{JHEP3} 
\usepackage{amsmath,amsthm,amsfonts,amssymb}

\def\be{\begin{equation}}\def\ee{\end{equation}}
\def\bea{\begin{eqnarray}}\def\eea{\end{eqnarray}}
\def\p{\partial}
\def\IC{\mathbb{C}}\def\IP{\mathbb{P}}
\def\IZ{\mathbb{Z}}
\def\eps{\epsilon}
\def\DL{\mathfrak{D}}

\def\qcor#1{\langle #1 \rangle}
\def\cx#1{{\cal#1}}\def\mx#1{{\mathfrak{#1}}}

\mathchardef\mm="2D
\def\Gr{\operatorname{Gr}}\def\td{{\tilde d}}
\def\tr{\operatorname{tr}}\def\Ws{{\hat W}}\def\cB{{\cal B}}
\def\KGr{K(\Gr)}
\title{Wilson loop algebras and quantum K-theory for Grassmannians}
\author{Hans Jockers$^1$, Peter Mayr$^2$, Urmi Ninad$^1$ and Alexander Tabler$^2$\\
$^1\,$Bethe Center for Theoretical Physics,\\
Physikalisches Institut, Universit\"at Bonn, 53115 Bonn, Germany\\
{\tt jockers@uni-bonn.de,urmi@th.physik.uni-bonn.de} \\ \\
$^2\,$Arnold Sommerfeld Center for Theoretical Physics,\\
Ludwig-Maximilians-Universit\"at,
80333 Munich, Germany\\
{\tt mayr@physik.uni-muenchen.de,tabler.alexander@physik.uni-muenchen.de} }
\abstract{We study the algebra of Wilson line operators in three-dimensional $\cx N=2$ supersymmetric $U(M)$ gauge theories with a Higgs phase related to a complex Grassmannian $\Gr(M,N)$, and its connection to K-theoretic Gromov--Witten invariants for $\Gr(M,N)$. For different Chern--Simons levels,  the Wilson loop algebra realizes either the quantum cohomology of $\Gr(M,N)$, isomorphic to the Verlinde algebra for $U(M)$, or the quantum K-theoretic ring of Schubert structure sheaves studied by mathematicians, or closely related algebras.}
\preprint{BONN--TH--2019--08\\LMU-ASC 53/19}

\begin{document}
\section{Introduction and outline}
The relation \cite{WittenVer}  between the Verlinde algebra \cite{VL88} and quantum cohomology of Grassmannians \cite{Vafa:1990mu,KI91,CV92} has been derived in ref.~\cite{Nekphd,KW13,NN17} by studying the small radius limit in an $S^1$-compactification of a 3d $\cx N=2$  supersymmetric gauge theory. This setup fits into the general framework of refs.~\cite{NekSha,NekSha2}, which connects 3d supersymmetric gauge theories to quantum K-theory, as opposed to quantum cohomology. An explicit 3d gauge theory/quantum K-theory correspondence between the $\cx N=2$ supersymmetric gauge theory and Givental's equivariant quantum K-theory \cite{Giv15all} was proposed and studied in ref.~\cite{JM}. For non-trivial Chern--Simons level, it involves the generalization to non-zero level defined in ref.~\cite{RZ18}. A concrete relation between quantum K-theoretic correlators and the Verlinde numbers was proposed in \cite{RZ18b}.

In this note we carefully reconsider the relation between the Verlinde algebra, quantum cohomology and quantum K-theory of Grassmannians by studying the algebra of 3d Wilson line operators. The 3d Wilson loop algebra depends on the choice of Chern--Simons (CS) levels $\kappa$ and the matter spectrum. In a certain window for $\kappa$, the dimension of this 3d algebra is fixed and equal to $\dim (H^*(\Gr(M,N)))$, but the quantum corrections to the structure constants of the ring of chiral operators associated with the loop algebra  depend on the values of $\kappa$. The family of quantum algebras for different $\kappa$ contains the Verlinde algebra and quantum cohomology, the ordinary quantum K-theory studied by Buch and Mihalcea~\cite{BM08}, and closely related algebras. The dependence on $\kappa$ drops out in the 2d scaling limit, where the quantum algebras become isomorphic to quantum cohomology, as predicted by the arguments of refs.~\cite{WittenVer, KW13}.

\section{Generalized $I$-functions}
In the following we determine the Wilson line algebra by mapping Wilson line operators in a representation $R$ of the 3d $\mathcal{N}=2$ $U(M)$ gauge theory to difference operators acting on Givental's $I$-function in quantum K-theory \cite{Giv15all}. The $I$-function is the generating function of K-theoretic correlators with at least one insertion, defined as an Euler number on an appropriate compactification of the moduli space of maps from a curve $\Sigma$ to the target space $X$. The basic difference operators $p_a$ shift the exponentiated, complexified K\"ahler parameters $Q_a$ as $p_a Q_a = q Q_a$, $a=1,\ldots, h^{1,1}(X)$. Here $q$ is the weight for the twisting of the world-volume geometry $C\times_q S^1$ of the 3d theory, where the Riemann surface $C$ is topologically a two-sphere $S^2$ or a disk $D^2$. These twisted world-volume geometries relate to the Omega background of refs.~\cite{Nekphd,NekWit}. In Givental's formalism, $Q_a$ represent Novikov variables and $q$ is the weight for the $\IC^*$ action in equivariant Gromov--Witten theory \cite{GivEGW}. A relation between Wilson line operators and difference operators has been first studied in refs.~\cite{Nekphd,Dimofte:2011py,BDP}. A difference module structure in the variables $Q_a$ in quantum K-theory has been described in ref.~\cite{GivTon}. For complete flag manifolds, the difference module had been related to the difference Toda lattice already in ref.~\cite{GivLee}. Such a connection between gauge theory and integrable systems holds much more generally, as shown in the context of Bethe/gauge correspondence~\cite{NekSha2}. 

\subsection{Generalized $I$-functions of $\Gr(M,N)$} 
The 3d gauge theory/quantum K-theory correspondence of ref.~\cite{JM} equates the $I$-function of permutation symmetric quantum K-theory of ref.~\cite{Giv15all} with the vortex sum of a supersymmetric partition function on $D^2\times_q S^1$. There are by now several methods to determine this partition function \cite{Gadde:2013wq,YS,BDP,DGP}. In this note we study the partition function for 3d $\mathcal{N}=2$ $U(M)$ gauge theory with $N$ chiral matter multiplets in the fundamental representation, which realizes (for $N>M$) the complex Grassmannian $\Gr(M,N)$ as its target space. The details of the computation of the partition function will be given elsewhere \cite{toappear}; the result is\footnote{To simplify some of the following formulas, a sign factor $(-1)^{M}$ has been absorbed in the definition of $Q$  relative to the conventions used in the mathematical literature.\label{fnsign}}
\bea\label{IGr}
I^{SQK}_{\Gr(M,N)}=c_0\, \sum_{0\leq d_a\in\IZ} (-Q)^{\sum\td_a} q^{CS(\td)}\frac {\prod_{b>a}^Mq^{\tfrac 12 \td_{ab}^2}\ (q^{\tfrac 12 \td_{ab}}-q^{-\tfrac 12 \td_{ab}})}
{\prod_{\alpha=1}^N\prod_{a=1}^M\prod_{r=1}^{d_a}(1-y_\alpha q^{r-\eps_a})}\, .
\eea
Here $c_0$ is a normalization factor chosen such that $ I^{SQK}_{\Gr(M,N)}|_{Q=0}=(1-q)$. The parameters $y_\alpha$ arise as real masses of the $N$ matter multiplets and define the equivariant $I$-function with respect to global flavor symmetry. Moreover $\td_{ab}=\td_a-\td_b$ and $\td_a = d_a-\eps_a$, where the parameters $\eps_a$ are related to the Chern roots $x_a$ of the dual $S^*$ of the tautological $U(M)$  bundle $S$ over $\Gr(M,N)$ by $q^{\eps_a} = e^{\beta x_a}$. Here $\beta$ is the radius of $S^1$, which will be set to $\beta =1$ in most of the following. For vanishing real masses $y_\alpha = 1$, the $I$-function takes values in the topological K-group $K(\Gr(M,N))$ ---  abbreviated in the following by $\KGr$.\footnote{More precisely, the $I$-function takes values in the equivariant K-group, and we obtain the K-group $K(\Gr(M,N))$ in the non-equivariant limit $y_\alpha=1$. In the following we mostly consider the non-equivariant limit.} Using the Chern isomorphism, a basis of $\dim(\KGr)={N\choose M}$ elements is provided by the Schur polynomials $\sigma_\mu(x^K)$ in the variables $x^K_a=1-e^{-x_a}$, where $\mu$ is a Young tableau in the set of tableaus fitting in a $M\times(N-M)$ box $\cB_N$. The expansion 
\be\label{Iexpansion}
I^{SQK}_{\Gr(M,N)}=\sum_{\mu \in \cB_N}  I_\mu(q,Q) \, \sigma_\mu(x^K) \ ,
\ee
represents the overlap of the groundstate of the 3d theory with boundary states defined by a 3d generalization of B-branes \cite{CV13,JM}. The choice of the variables $x^K_a$ is motivated by the 2d limit $\beta\to 0$, where the basis $\{\sigma_\mu(x^K)\}$ reduces to the cohomological basis dual to the Schubert cycles, $\sigma_\mu(x^K)\, {\buildrel \beta \to 0 \over \longrightarrow}\, \beta^{|\mu|}(\sigma_\mu(x)+O(\beta))$.\footnote{The connection of the K-theory basis $\{\sigma_\mu(x^K)\}$ to the basis of  Grothendieck polynomials defined in ref.~\cite{MR686357} will be discussed in sect.~\ref{sec:examples}.}

The {\it effective} CS levels in the 3d gauge theory are encoded in the term\footnote{For a matrix $\sigma=\textrm{diag}(\sigma_1,\ldots,\sigma_M)$ the trace symbols are defined as $\tr_{U(M)}(\sigma^2)=\sum_a\sigma_a^2$,  $\tr_{U(1)}(\sigma^2)=\frac 1 M (\sum_a\sigma_a)^2$, $\tr_{SU(M)}(\sigma^2)=\tr_{U(M)}(\sigma^2)-\tr_{U(1)}(\sigma^2)$ and $\tr_R(\sigma)=\sum_a\sigma_a$.}
\be
q^{CS(\td)} = q^{\frac 12\, \hat \kappa_S\,\tr_{SU(M)}(\td^2) + \frac 12 \,\hat \kappa_A \,\tr_{U(1)}(\td^2)+ \hat \kappa_R \,\tr_R(\td)} \ .
\ee
The parameters $\hat \kappa_S$ and $\hat \kappa_A$ specify the levels for the $SU(M)$ and $U(1)$ subgroups of the gauge group $U(M)$, respectively, while $\hat\kappa_R$ is a level for the mixed gauge/R-symmetry CS term. The effective CS levels $\hat\kappa_i$ fulfill\footnote{The quantization relation linking $\hat\kappa_S$ and $\hat\kappa_A$ is a consequence of the global structure of the Lie group $U(M) \simeq \left( SU(M) \times U(1) \right) / \mathbb{Z}_M$.}
\be
   \hat\kappa_S \in \mathbb{Z} \ , \qquad \frac{\hat\kappa_S - \hat\kappa_A}M \in \mathbb{Z} \ , \quad 2 \hat\kappa_R \in \mathbb{Z} \ ,  
\ee
and the relation to the bare bulk CS terms $\kappa_i$ in the Lagrangian are \cite{toappear}
\be\label{bulkCS}\textstyle
\hat\kappa_S = \kappa_S-M+\frac N2\ ,\quad
\hat\kappa_A = \kappa_A+\frac N2\ ,\quad
\hat\kappa_R = \kappa_R+\frac N4 \ .
\ee
Following the arguments of ref.~\cite{JM}, the  level zero quantum K-theory of ref.~\cite{Giv15all} corresponds to zero effective CS levels $\hat\kappa_{i}=0$. For other levels $\hat \kappa_i$, one obtains a three parameter family of generalized $I$-functions. One can check that on the one-parameter slices $(\hat \kappa_S,\hat \kappa_A,\hat \kappa_R)=(\ell_\square,\ell_\square,-\ell_\square/2)$ and $(\hat \kappa_S,\hat \kappa_A,\hat\kappa_R)=(0,M\ell_{\operatorname{det}},-\ell_{\operatorname{det}}/2)$ the above result respectively reproduces the $I$-functions  at level $\ell_\square$ in the fundamental representation of $U(M)$ and at level $\ell_{\operatorname{det}}$ in the determinantal representation of $U(M)$ in quantum K-theory with level structure, as derived in ref.~\cite{Wen} (see also ref.~\cite{RZ18b}).\footnote{Different K-theoretic $I$-functions for complex Grassmannians 
have also appeared in refs.~\cite{Tai,Woodward:2018rab}, however we did not find a suitable choice of CS terms~\eqref{bulkCS} such that these match with the field theory result~\eqref{IGr}.} 

To study the action of Wilson line operators in a representation $R$ of $U(M)$ we will also need an abelianized version of the path integral for the maximal torus $U(1)^M\subset U(M)$.\footnote{Abelianization of Grassmannian sigma models has been used many times before in physics and mathematics; for some literature closely related to the present context see refs.~\cite{CV92,HV,Martin,Wen}.} To this end we consider a modified sum $\hat I(Q_a)$ depending on $M$ weights $Q_a$,
\be
\hat I(Q_a) =c_0\, \sum_{\vec d \in\IZ_{\ge 0}^M}  c_{\vec d}\ (-Q)^{\vec d}\ , \qquad (-Q)^{\vec d}=\prod_a(-Q_a)^{\td_a} \ ,
\ee
where the coefficients $c_{\vec d}$ are defined by the requirement $ \hat I(Q_a)  \big|_{Q_a = Q} = I^{SQK}_{\Gr(M,N)}(Q)$.
The sum $\hat I$ satisfies the relation:
\be\label{defIh}
\hat I=\Delta\, \tilde I\ ,\qquad \Delta = \prod_{b>a} (p_a-p_b)\ ,
\ee
which can be viewed as a K-theoretic version of the Hori--Vafa formula \cite{HV}. Here $p_a=q^{\theta_a}$, $\theta_a=Q_a\p_{Q_a}$, and
\bea\label{defIt}\textstyle
\tilde I(Q_a)&=&c_0\, \sum_{\vec d \in\IZ_{\ge 0}^M} q^{\gamma \sum_{b>a}\td_a\td_b}\prod_a I^{d_a,\eps_a,\alpha,\beta}_{\IP^{N-1}}(-Q_a )\, .
\eea
The constants are
\be\label{const}
\alpha = \hat \kappa_S+ \Delta_\kappa+M \ , \quad 
\beta = \hat \kappa_R-\frac 12 (M-1) \ ,  \quad 
\gamma =\frac{\hat \kappa_A-\hat\kappa_S}{M}-1=\Delta_\kappa\ ,  
\ee
with $\Delta_\kappa=\tfrac{\kappa_A-\kappa_S}M$. The last factor in eq.~\eqref{defIt} involves the coefficients of a generalized $I$-function $I^{\alpha,\beta}_{\IP^{N-1}}=\sum_{d\geq 0} I^{d,\eps,\alpha,\beta}_{\IP^{N-1}}(Q)$ for  $\IP^{N-1}$  obtained from a $U(1)$ gauge theory with CS levels $(\alpha,\beta)$, which reads
\be\label{IPN}
I^{d,\eps,\alpha,\beta}_{\IP^{N-1}}(Q)= \frac{Q^{\td}q^{\frac{\alpha}2  \td^2+\beta \td}}{\prod_{r=1}^{d}(1-q^{r-\eps})^N}\ .
\ee
The cross terms $\sim q^{\gamma\, \td_a\td_b}$ in eq.~\eqref{defIt} obstruct a complete factorization into a product of $M$ $\IP^{N-1}$ factors, except for $\Delta_\kappa=0$. This condition is however incompatible with the canonical levels $\hat \kappa_i=0$ relevant for the level zero quantum K-theory.\\ 

\subsection{A window for the Chern--Simons levels}
The 3d gauge theory/quantum K-theory correspondence equates the vortex sum of the 3d gauge theory with the $I$-function for the permutation symmetric K-theory of refs.~\cite{Giv15all,RZ18}. That is, the $I$-function $I_{\Gr(N,M)}^{SQK}(\delta t)$ in eq.~\eqref{IGr} has in general a non-zero permutation symmetric input $\delta t$, determined by the expansion
\be\textstyle
I^{SQK}_{\Gr(M,N)}(\delta t)=(1-q)+\delta t +I_{\textrm corr}(\delta t)\ ,\qquad \delta \in \cx K_+\ ,\quad I_{\textrm corr}\in \cx K_-\ ,
\ee
which separates the perturbations $\delta t$ from the correlator part $I_{corr}(\delta t)$.
Here $\mathcal{K} = \KGr \otimes \mathbb{C}(q,q^{-1})\otimes\mathbb{C}[[Q]]$ is the symplectic loop space,  and $\mathcal{K}_\pm$ the Lagrangian subspaces defined as \cite{GivTon,GivER,RZ18}
\be
\begin{aligned}
  \mathcal{K}_+&=\KGr \otimes \mathbb{C}[q,q^{-1}]\otimes\mathbb{C}[[Q]]  \ ,  \\
  \mathcal{K}_- &=\KGr \otimes  \left\{\,r(q) \in R(q) \,\middle|\, \text{$r(0)\ne\infty$ and $r(\infty)=0$} \right\} \otimes\mathbb{C}[[Q]] \ , 
\end{aligned}  
\ee
where $R(q)$ denotes the field of rational functions in the variable $q$.
The permutation symmetric input corresponds to a 3d partition function perturbed by multi-trace operators \cite{JM}. To compute the chiral ring one wants to study the family of 3d gauge theories perturbed only by single trace operators, corresponding to the ordinary quantum K-theory. Hence, one has to shift the multi-trace perturbations to zero, and subsequently perturb in the single trace directions, $I^{SQK}(\delta t)\to I^{SQK}(0)=I^{QK}(0)\to I^{QK}(t_{\textrm{single trace}})$. 

There is a certain window of CS levels, for which the $I$-function $\eqref{IGr}$ is already at zero input, $\delta t=0$, and coincides with the unperturbed $I$-function $I^{QK}_{\Gr(M,N)}(0)$ of ordinary quantum K-theory. Inspection of eq.~\eqref{IPN} shows that this requires the CS levels to lie in the window
\be\label{levwin}
H_\kappa:\qquad |\kappa_S+\Delta_\kappa|\leq \frac N 2\ .
\ee
As discussed in the next section, the same condition assures that the $\mathbb{Z}[[Q]]$-module of Wilson loops has $\binom{N}{M}$ generators  --- in the following referred to as the dimension of the Wilson loop algebra --- which is the number of generators of the K-theory group $\KGr$. There are extra operators outside the window $H_\kappa$.

\section{Algebra of difference operators \label{sec:WLA}}
\subsection{Wilson loops}
In the 3d gauge theory one can consider more generally expectation values $\qcor{W_R}$ of Wilson loop operators $W_R=\tr_R\textrm{P\,exp}(i\int_{S^1}(A_\theta-i\sigma)d\theta)$ wrapping $S^1$, where $\sigma$ is the real scalar field in the 3d vector multiplet and $R$ is a representation of the gauge group.\footnote{See e.g. refs.~\cite{KW09, KW13, CV13,YS,NN17} for studies of Wilson loops in Chern--Simons matter theories.} The insertion of a charge minus one Wilson line in the $a$-th $U(1)$ factor in the path integral produces an extra factor $q^{\td_a}$ in the sector with vortex number $d_a$. The vacuum expectation value of a $U(1)^M$ Wilson line $W_{\vec n}=\prod_a W_a^{n_a}$ is then (up to a normalization factor) equal to a vortex sum
\be\label{WLu}
\hat I_{\vec n} =c_0\, \Delta\,  \sum_{\vec d\in\IZ_{\ge 0}^M}  (-Q)^{\vec d} \tilde c_{\vec d}\, \prod_a q^{\td_an_a}  \ ,
\ee
where $\Delta$ is the operator defined in eq.~\eqref{defIh} and $\tilde c_{\vec d}$ are the coefficients of the $I$-function $\tilde I=c_0\, \sum_{d_a\geq 0} (-Q)^{\vec d}\tilde c_{\vec d}$ defined in eq.~\eqref{defIt}.

The extra factor $ \prod_a q^{\td_an_a}$ can be generated by acting with a difference operator $\hat I$. Defining $\DL_{\vec n}(p)=\prod_a p_a^{n_a}$, with $p_a = q^{\theta_a}$, $\theta_a = Q_a\p_{Q_a}$, one has
\be
\hat I_{\vec n}=\DL_{\vec n} (p)\, \Delta \tilde I \ .
\ee
Classically, a product of two Wilson line operators $W_{\vec n}$ and $W_{\vec m}$ corresponds to the insertion of $\prod_a q^{\td_a(m_a+n_a)}$ in the sum, or the action with $\DL_{\vec n}\cdot \DL_{\vec m}=\DL_{\vec m+\vec n}$ on $\hat I$. In the quantum theory, there are non-trivial relations between the Wilson lines for different $\vec n$. To this end, one notices that the sum $\tilde I$ satisfies the  following difference equations:
\bea \label{DiffIt}
\mx L_a \,  \tilde I &=& O(\eps_a^N)\, ,\quad  a=1,\ldots,M\, ,\nonumber\\[2mm]
\mx L_a&=& (1-p_a)^N\, +Q_a q^{\frac{\alpha}2 + \beta}p_a^\alpha\prod_{b\neq a} p_b^\gamma\, .
\eea
These equations are  inherited from the difference equations of the $\IP^{N-1}$ factors, taking into account the additional CS terms. The $O(\eps_a^N)$ terms are set to zero in $H^*(Gr(M,N))$. The difference equations allow to replace a difference operator $p_a^D$ acting on $\tilde I$ by an operator of lower degree in $p_a$. Here $D$ is the degree in $p_a$ of the polynomial obtained by clearing denominators in the operator $\mx L_a$:
\be
D=\begin{cases} N & \text{for } |\kappa_S+\Delta_\kappa|\leq \frac N2\, ,\\|\kappa_S+\Delta_\kappa|+\frac N2& \text{for } |\kappa_S+\Delta_\kappa|>\frac N2\, .
\end{cases}
\ee
This reduces the algebra of difference operators acting on $\tilde I$ to $\DL_{\vec n}(p)/(p_a^D)$. Taking into account the extra operator $\Delta$ in $\hat I$, which has degree $M-1$ in each variable $p_a$, the algebra of difference operators acting on $\hat I$ is generated as a $\mathbb{Z}[[Q]]$-module by the operators $\{\DL_{\vec n},\, 0\leq n_a\leq D-M\}$. In view of the relations \eqref{DiffIt}, it is useful to also define shifted Wilson line operators $\Ws_a=1-W_a$, corresponding to the action of the difference operator $\delta_a = 1-p_a$. 

In the $U(M)$ theory, one has to restrict to permutation symmetric combinations of $U(1)^M$ Wilson lines, which can be represented by Schur polynomials $\sigma_\mu$ (labeled by Young tableaus $\mu$) either in the operators $\delta_a$ or the operators $p_a$, i.e., 
\be
W_\mu \leadsto \DL_\mu(p) \ \ \text{with}\ \  \DL_\mu = \sigma_\mu(p_a)\ , \qquad
\Ws_\mu \leadsto \hat\DL_\mu(\delta) \ \ \text{with}\ \ \hat\DL_\mu = \sigma_\mu(\delta_a)\ .
\ee 
For explicitness we focus now on the operators $\hat\DL_\mu$, which satisfy the algebra 
\be\label{WLalg1}
\hat\DL_\mu* \hat\DL_\nu = \sum_{\lambda} C_{\mu\nu}^\lambda \hat\DL_\lambda\, ,
\ee
where the structure constants $C_{\mu\nu}^\lambda $ are the Littlewood--Richardson coefficients. Classically the ideal of relations is $\delta_a^{D-M+1}=0$ , i.e., $\qcor{\Ws_\lambda}=0$ for representations $\lambda$ not fitting into a $M\times (D-M)$ box $\cB_{D}$. This yields the number of representation-theoretic independent Wilson lines $\hat W_\lambda$, which is given by
\be
\tr (-)^F = {D\choose M}\ .
\ee
For $\Delta_\kappa=0$ one recovers the 3d Witten index computed already in refs.~\cite{BZ,ClK}.\footnote{This UV computation of the 3d Witten index agrees with an IR computation along the lines of ref.~\cite{IS}, see ref.~\cite{toappear}.} After the reduction by the ideal \eqref{DiffIt}, the structure constants depend on $Q,q$ and the CS levels, namely
\be
\hat\DL_\mu* \hat\DL_\nu =\sum_{\lambda\in\cB_D} c_{\mu\nu}^\lambda(Q,q,\kappa_i)\, \hat\DL_\lambda\ .
\ee
Here the structure constants $c_{\mu\nu}^\lambda(Q,q,\kappa_i)$ are obtained from the action of the $\DL_\mu$ on $\hat I$ and taking  the $Q_a=Q$ limit. One has $\left.c_{\mu\nu}^\lambda\right|_{Q=0}=C_{\mu\nu}^\lambda$. The reduced algebra of difference operators is {\it not} equal to the Wilson loop algebra, due to extra terms generated by the action of a difference operator on the $Q$-dependent terms in the relations \eqref{DiffIt}. Since
\be
\delta(f(Q)\hat I)=(\delta f)\hat I+f(Qq)\delta\hat I = f(Q)\delta \hat I + O(1-q)\ ,
\ee
 these terms are necessarily of $O(1-q)$ and can be eliminated by setting $q=1$. The algebra of Wilson line operators is therefore 
\be\label{WLalg}
\Ws_\mu* \Ws_\nu = \sum_{\lambda\in \cB_D} C_{\mu\nu}^\lambda(Q,\kappa_i)\,  \Ws_\lambda\ ,
\ee
with $C_{\mu\nu}^\lambda(Q,\kappa_i)= \left. c_{\mu\nu}^\lambda(Q,q,\kappa_i)\right|_{q=1}$. 

The equations \eqref{DiffIt} are related to the vacuum equations of ref.~\cite{NekSha2}, by first replacing $q^{\theta_a}$ by a commuting variable $p_a$ and subsequently taking the $q\to 1$ limit. The discussion is also similar to the treatment in ref.~\cite{KW13}, where the algebra of Wilson loops has been computed from an ideal of relations for a $S^3$-partition function.\footnote{See also refs.~\cite{BZ,ClK}.} It has been argued there, that the algebra should be independent of the manifold, on which the 3d theory is compactified. A comparison of the above results with those in ref.~\cite{KW13} will be made below.

\subsection{Diff. equations for the $I$-function \label{sec:Diffeq}}
As discussed around eq.~\eqref{levwin}, the permutation symmetric $I$-function \eqref{IGr} coincides with the unperturbed $I$-function of the ordinary quantum K-theory inside the level window $H_\kappa$. The latter can be perturbed by operators $\Phi_i\in \KGr$, introducing  a dependence on ${N \choose M}$ additional variables $t_i$. The correlators of the perturbed theory are captured by the Givental $J$-function\footnote{The $I$-function studied before, and the $J$-function below, are in general connected by a non-trivial transformation, the 3d mirror map. At zero perturbation, this transformation is trivial for $\Gr(M,N)$ with canonical CS levels, i.e., $J(0)=I(0)$. For non-zero level, the correlators and the inner product in the following two equations have to be defined as in ref.~\cite{RZ18}.}
\be
J(t) = (1-q)+t + \sum_{\beta, i,n\geq 0} \frac{Q^\beta}{n!}\qcor{\frac{\Phi_i}{1-qL},t^n}_{n+1,\beta}\,  \Phi^i \ .
\ee
The sum over $\beta$ runs over the degrees of rational curves and $n$ counts the number of insertions of the perturbations $t=\sum_i t_i\Phi_i\in\cx K_+$, where $\{\Phi_i\}$, $i=0,\ldots, {N\choose M}-1$ is a basis of $\KGr$. The elements $\Phi^i=\sum \chi^{ij}\Phi_j$ of the dual basis are defined with respect to the inner product
\be\label{ip}
(\Phi_i,\Phi_j)=\int_{\Gr(M,N)} \!\!\! \textrm{td}(\Gr(M,N)) \,\Phi_i\Phi_j=:\chi_{ij}\  .
\ee
The correlator part lies in $\cx K_-$ and is defined as a holomorphic Euler characteristic on the moduli space of stable maps \cite{GivWDVV,Lee:2001mb}.
 
The $J$-function of $\Gr(M,N)$ relates to a fundamental solution $T\in \textrm{End}(\KGr)$ of the flatness equations \cite{GivWDVV,IMT}
\be\label{fleq}
\begin{aligned}
(1-q) \p_{t_i} T &= T\Phi_i*\, , \qquad  &&i=0,\ldots,{N\choose M}-1\, ,\\
q^{\theta}T &= T Aq^{\theta}\, , &&\theta = Q \partial_Q \ ,  
\end{aligned}
\ee
where $*$ denotes the quantum product, $A\in \textrm{End}(\KGr)$ and we suppress the $Q,t,q$-dependence of the endomorphisms. In terms of $T$ one has 
\be
J(t)=(1-q)T\Phi_0=(1-q)\, \sum_i(T\Phi_0,\Phi_i)\, \Phi^i\, ,
\ee
where $\Phi_0=1$ denotes the unit.
The systems of first order equations \eqref{fleq} imply differential equations for the $J$-function. For concreteness and simplicity we consider the small deformation space spanned by the element $\Phi_1=1-e^{c_1(S)}$ and suppress the index on $t$. 
Suppose there is a relation in the quantum K-theoretic algebra of the form 
\be\label{nablarel}
\sum_{k=0}^d a_{k}\nabla^{k} \Phi_0 = 0\, ,\qquad \nabla=(1-q) \p_{t} + \Phi_1*\, ,
\ee
for some integer $d\leq {N \choose M}$, where $a_k$ are some $(Q,q)$-dependent coefficients. Then
\be
0=(T\sum_{k} a_{k}\nabla^{k} \Phi_0,\Phi_i)= \sum_{k} a_{k}\p^k_t(T\Phi_0,\Phi_i)\, ,
\ee
where the second step uses that the basis elements $\Phi_i$ are constant. This 
implies the differential equation 
\be\textstyle
\sum_{k=0}^d a_{k} D^k J =0\  , \qquad D=\p_t\, .
\ee
Similarly, a relation $\sum_{k} b_{k}(Aq^{\theta})^k \Phi_0 =0$ in the quantum algebra implies the difference equation 
\be\textstyle\label{gendiffeq}
\sum_{k} b_{k} p^k J =0\ ,\qquad \ \ \ p=q^\theta\, .
\ee
In this way, relations in the quantum K-theory algebra translate to quantum differential and difference equations for the $J$-function and vice versa.

\section{Selected examples \label{sec:examples}}
\subsection{Quantum cohomology and factorized case}
\noindent {\it Factorized case}\\
The simplest subset of theories arises for the choice of CS levels with $\Delta_\kappa=0$, i.e., equal {\it bare} CS levels in the $SU(N)$ and $U(1)$ factors, but unequal effective levels, $\hat \kappa_S\neq \hat \kappa_A$. In this case, the relations in eq.~\eqref{DiffIt} factorize into $M$ separate equations 
\be
\delta_a^N = -Q_aq^{\frac\alpha 2 +\beta}p_a^\alpha\, .
\ee
These quantum relations are equivalent to those obtained in ref.~\cite{KW13} from a $S^3$ partition function.

\noindent {\it A K-theory algebra isomorphic to quantum cohomology/Verlinde algebra}\\
In the window $|\kappa_S+\Delta_\kappa|\leq \frac N2$, the dimension of the Wilson line algebra equals $\dim(\KGr)={N\choose M}$, and the algebra of Wilson lines can be isomorphic to the quantum cohomology ring of $\Gr(M,N)$, which, furthermore, is isomorphic to the Verlinde algebra of the gauged Wess--Zumino--Witten model $U(M)/U(M)$ at level $(N-M)$~\cite{WittenVer}. This happens for the special case $\alpha=\beta=\gamma=0$, where the difference equation becomes 
\be (1-p_a)^N = -Q_a\ .\ee This is, up to our sign convention,  the same relation as in quantum cohomology \cite{Vafa:1990mu,KI91,CV92}, and thus will give an isomorphic algebra after projecting to $U(M)$. The operators generating these algebras are 3d Wilson lines $\Ws_\mu$ in the quantum K-theoretic ring or the corresponding operators related to the cohomology classes $[\sigma_\mu]\in H^*(\Gr(M,N))$ in the 2d case \cite{NN17}.\\

\noindent{\it 2d limit}\\
In the 3d gauge theory compactified on an $S^1$ of radius $\beta$, one can consider a simple small radius limit, in which the quantum K-theoretic $I$-function reduces to the cohomological $I$-function. To this end one sets $q=e^{-\hbar \beta}$, and rescales the operators and the FI parameter by factors of $\beta$ \cite{JM}. For any CS levels, the 2d limit of the relations \eqref{DiffIt}  is
\be\label{2dlim}
(\beta \theta_a)^N=-Q_a\, ,
\ee
which is again the relation in quantum cohomology, after a renormalization $Q_a\to Q_ae^{N\ln\beta}$. The different K-theoretic algebras of Wilson loops distinguished by different CS levels therefore reduce all to the quantum cohomology algebra, in agreement with the arguments of refs.~\cite{WittenVer,KW13}.

\subsection{Chiral ring at level zero}
The canonical case from the point of quantum K-theory is however at zero effective CS level, related the ordinary K-theory of refs.~\cite{GivWDVV,Lee:2001mb}. In the following we study the chiral ring from the algebra of difference operators and show that the result agrees with the results in the mathematical literature on the quantum product of Schubert structure sheaves \cite{BM08} .

The canonical case lies inside the level window $H_\kappa$ if $N>M$, which is required for the theory to have a  supersymmetric vacuum, see e.g., ref.~\cite{KW13}. The chiral ring can therefore be studied directly from the $I$-function \eqref{IGr} at zero input. The difference equations \eqref{DiffIt} take the form 
\be\label{Ican}
\delta^N_a =- Q_a \frac{p_a^M}{\prod_{b=1}^M p_b}\ , \quad a=1,\ldots, M\, .
\ee
The appearance of a denominator on the r.h.s. is harmless, as the operators $p_a$ are invertible. Due to the non-factorized form, the general polynomial reduction defined by the ideal \eqref{Ican} is not entirely trivial, but straightforward to perform for given values of $M$ and $N$. 

As an illustrative example we discuss the first non-trivial case $\Gr(2,4)$. Performing the polynomial reduction of the difference operators one obtains, up to order $O(1-q)$ terms, the multiplication table \def\Wh{\sigma}
\be\label{tabg24}
\begin{small}
\hbox{
\vbox{\offinterlineskip
\halign{\strut\hskip-0.2cm$#$\hfil&~~=~~$#$\,,\hfil\hskip0.5cm&$#$\hfil&~~=~~$#$\hfil\cr
\Wh_1*\Wh_1&\Wh_2+\Wh_{1,1}&\Wh_2*\Wh_{2,2}&Q(\sigma_{1,1}+\sigma_{2,1}-\sigma_{2,2})+Q^2\,,\cr
\Wh_1*\Wh_2&\Wh_{2,1}+Q(2+\rho_1)&\Wh_{1,1}*\Wh_{1,1}&\Wh_{2,2}\,,\cr
\Wh_1*\Wh_{1,1}&\Wh_{2,1}&\Wh_{1,1}*\Wh_{2,1}&Q\rho_1\,,\cr
\Wh_1*\Wh_{2,1}&\Wh_{2,2}+Q(1+\rho_1)&\Wh_{1,1}*\Wh_{2,2}&Q(\Wh_2-\Wh_{2,1})\,,\cr
\Wh_1*\Wh_{2,2}&Q\rho_1&\Wh_{2,1}*\Wh_{2,1}&Q(\Wh_2+\Wh_{1,1}-\Wh_{2,2})+Q^2\,,\cr
\Wh_2*\Wh_{2}&\Wh_{2,2}+Q(\sigma_1+\rho_2)+Q^2&\Wh_{2,1}*\Wh_{2,2}&Q(\Wh_{2,1}-\Wh_{2,2})+Q^2\,,\cr
\Wh_2*\Wh_{1,1}&Q(1+\rho_1)&\Wh_{2,2}*\Wh_{2,2}&Q^2\,.\cr
\Wh_2*\Wh_{2,1}&Q\rho_2+Q^2\cr
}}}
\end{small}\ee
Here $\sigma_\mu$ stands for either $\Ws_\mu$ or $\hat\DL_\mu(\delta)$ and we used the 
abbreviations $\rho_1=\Wh_1+\Wh_2-\Wh_{1,1}-\Wh_{2,1}$, $\rho_2=\Wh_1+\Wh_2+\Wh_{1,1}-\Wh_{2,2}$. 

In the classical limit $Q=0$, the difference operators $\sigma_\mu(\delta)$ become Schur polynomials $\sigma_\mu(x^K_a)$ in the variables $x^K_a=1-e^{-x_a}$. The polynomials $\sigma_{\mu}(x^K)$ represent Chern characters of a basis $\{\phi_\mu\}$ for $\KGr$. The pairing $\chi(\phi_\mu,\phi_\nu)$ on the basis $\{\phi_\mu\}$ has already determinant one, but is non-minimal in the sense that the entries are not all 0 or 1. The polynomials $\sigma_{\mu}(x^K)$ are related by a linear transformation of determinant one to the Grothendieck polynomials $\cx O_\mu(x^K)$ of ref.~\cite{MR686357}:
\bea\label{baschange}
\cx O_1  &=& \sigma_1 -\sigma_{1,1} \, ,\qquad \quad
\cx O_2  = \sigma_2 -\sigma_{2,1}\,  ,\qquad 
\cx O_{1,1}  = \sigma_{1,1}\, ,\qquad \nonumber\\
\cx O_{2,1}  &=& \sigma_{2,1} - \sigma_{2,2}\,  ,\qquad 
\cx O_{2,2}  = \sigma_{2,2}\,  .\qquad 
\eea
The inner product \eqref{ip} on the new basis  is 
\be 
\chi(\cx O_\mu,\cx O_\nu) = \begin{pmatrix}
 1 & 1 & 1 & 1 & 1 & 1 \\[-1.5mm]
 1 & 1 & 1 & 1 & 1 & 0 \\[-1.5mm]
 1 & 1 & 1 & 0 & 0 & 0 \\[-1.5mm]
 1 & 1 & 0 & 1 & 0 & 0 \\[-1.5mm]
 1 & 1 & 0 & 0 & 0 & 0 \\[-1.5mm]
 1 & 0 & 0 & 0 & 0 & 0 
\end{pmatrix},\qquad \det(\chi)=1\ .
\ee
The $\cx O_\mu$  are the Chern characters of the the structure sheaves of the Schubert cycles. After the basis change, we obtain for the quantum multiplication of the structure sheaves 
\be\label{Tab24}\begin{small}
\begin{array}{c|ccccc}
  *& \cx O_{1} & \cx O_{2} & \cx O_{1,1} & \cx O_{2,1} & \cx O_{2,2} \\
\noalign{\hrule}
 \cx O_{1} & \cx O_{2}+\cx O_{1,1}-\cx O_{2,1} &-  & - & - & - \\
 \cx O_{2} & \cx O_{2,1} & \cx O_{2,2} & - & - & - \\
 \cx O_{1,1} & \cx O_{2,1} & Q & \cx O_{2,2} & - & - \\
 \cx O_{2,1} & \cx O_{2,2} + Q\,  (1-\cx O_{1})& Q\,  \cx O_{1} & Q\,  \cx O_{1} & Q\, 
   (\cx O_{2}+\cx O_{1,1}-\cx O_{2,1}) & - \\
 \cx O_{2,2} & Q\,  \cx O_{1} & Q\,  \cx O_{1,1} & Q\,  \cx O_{2} & Q\,  \cx O_{2,1} & Q^2 \\
\end{array}
\end{small}\ee
These multiplications agree with the result of ref.~\cite{BM08}, which has been obtained by quite different methods. 

At zero deformations $t=0$, as detailed in ref.~\cite{toappear}, the K-theoretic quantum product can be also obtained from the critical locus of the 1-loop effective $\mathcal{N}=2$ twisted superpotential $\widetilde W$ \cite{NekSha2}. In the example, there are two relations obtained from $e^{d \widetilde W} =1$, which can be written as
\be\label{Gen24}
r_1=\cx O_1^3-2\cx O_1\cx O_{1,1}+\cx O_1^2\cx O_{1,1} \ ,\qquad
r_2=\cx O_1^2\cx O_{1,1}-\cx O_{1,1}^2+\cx O_1\cx O_{1,1}^2-Q\ .
\ee
One can check that, similiar to the cohomological case \cite{Vafa:1990mu,KI91,CV92}, the K-theoretic quantum product~\eqref{Tab24} is isomorphic to a polynomial ring in two generators $\cx O_1,\cx O_{1,1}$, divided by the ideal of relations $r_a=0$
\be
\mathbb{Z}[\cx O_1,\cx O_{1,1},Q]/(r_1,r_2)\ .
\ee

The Grassmannians $\Gr(M,N)$ and $\Gr(N-M,N)$ are related by the well-known classical geometric duality, and this duality persists at the level of quantum K-theory. Under the duality a Young tableau is exchanged with its transposed Young tableau. In the self-dual example $\Gr(2,4)$, the duality in the quantum theory is reflected in the symmetry of the quantum multiplication \eqref{Tab24} under the exchange of $\cx O_2$ and $\cx O_{1,1}$. In the app.~\ref{sec:appG25} we discuss the less trivial example of $\Gr(2,5)\simeq \Gr(3,5)$ of dual quantum K-theories and its relation to Seiberg like dualities of ref.~\cite{BCC}.

\subsection{Difference equations at level zero} \label{sec:DiffeqSec}
To determine the difference equations satisfied by $I_{\Gr(M,N)}(0)$ in \eqref{IGr}, it suffices to know the quantum product at zero deformation  computed above. It is straightforward to show, that at $t=0$, the action of $A$ in eq.~\eqref{fleq} is equal to the quantum product with $\Phi_1$, i.e. $A|_{t=0}=(\Phi_1*)|_{t=0}$. Computing the linear relations between suitable powers of $(Aq^\theta)^k\Phi_0$, we find the difference operators annihilating the $I$-function $I_{\Gr(M,N)}(0)$ listed in Table~\ref{tab:diffop}.
\begin{table}[h]\begin{center}
\hbox{
\vbox{\offinterlineskip
\halign{\strut~$#$~\vrule&~~$#$~~\hfil\vrule&~~$#$~\hfil\cr
\!\!\!\!(M,N):&\textrm{3d difference operator}&\textrm{2d limit}\cr
\noalign{\hrule}
(2,3):& \delta^3-Q&\theta^3-\tilde Q\cr
(2,4):& \delta ^5+Q (p q+1) (p^2 q-1 )&\theta^5-2 (2 \theta+1) \tilde Q\cr
(3,4):& \delta ^4+Q &\theta^4+\tilde Q\cr
(2,5):& \delta^7 (-1 + \delta + q)^3+O(Q)&(\theta-1)^3 \theta^7-\left(11 \theta^2+11 \theta+3\right) \theta^3 \tilde Q-\tilde Q^2\cr
(2,6):&\delta ^9 (\delta +q-1)^5 &(\theta-2)\big[- \theta^5 (1 + 2 \theta) (4 + 13 \theta + 13 \theta^2) \tilde Q \cr
&\quad\ \left(\delta +q^2-1\right)+O(Q)&\quad+(-1 + \theta)^5 \theta^9 - 
 3 (2 + 3 \theta) (4 + 3 \theta) \tilde Q^2\big]\cr
}}}
\end{center}\ \\[-60pt]
\caption{Difference operators for low values of $(M,N)$ and their 2d limits\label{tab:diffop}}
\end{table}
\\The difference operators for $\Gr(2,5)$ and $\Gr(2,6)$ are of degree 4 and 8 in $Q$ and the complete expressions are  lengthy, see e.g. eq.~\eqref{dop25}. On the r.h.s. we have included the 2d limit of the difference operator described around eq.\eqref{2dlim}, which gives a differential operator in $\tilde Q$. Here $\tilde Q = Q\beta^{-N}$ is the renormalized parameter with  $\beta$ the $S^1$ radius. These limits agree with the differential operators of ref.~\cite{Batetal} that annihilate the cohomological $I$-function. 

\subsection{Perturbed theory for $\Gr(2,4)$}
The above results for the $I$-function and the chiral ring hold at zero perturbation $t=0$. Methods to reconstruct the $I$-function at non-zero $t$ from the $I$-function at $t=0$ have been described in refs.~\cite{IMT,GivER,Giv15all}, using the difference module structure in quantum K-theory, first established in ref.~\cite{GivTon} for the ordinary theory. 

These reconstruction theorems use the difference operators \eqref{fleq} to deform the $I$-function and are therefore limited to the directions in $K(X)$ generated by repeated action of $A$ on the unit $\Phi_0$. In the case of the Grassmannian $\Gr(M,N)$, this reconstructs the subspace of the deformations $t=\sum_i t_i(\cx O_1)^i$ generated by the difference operator associated with the line bundle $e^{c_1(S)}=1-\cx O_1$. 

From the view point of the 3d gauge theory, the full deformation space can be accessed by adding single trace deformations in any representation of the gauge group \cite{JM}, and this is not limited to the subspace generated by powers of $\cx O_1$. It suffices to know the action of Wilson line operators $\hat W_\mu$ on the $I$-function for any representation~$\mu$. This action has been already reconstructed in sect.~\ref{sec:WLA} in terms of the difference operators $\hat \DL_\mu$ acting on the generalized $I$-function $\hat I(Q_a)$ depending on $M$ Novikov variables $Q_a$, and setting $Q_a=Q$ at the end. 

The general reconstruction for $\Gr(M,N)$ will be discussed in ref.~\cite{toappear}. Here we briefly discuss the simplest case of the level zero quantum K-theory for $\Gr(2,4)$ at non-zero $t$, where one can use duality to reconstruct the $I$-function from the reconstruction methods of refs.~\cite{IMT,GivER}. Classically, the Grassmannian $\Gr(2,4)$ can be described via the Pl\"ucker embedding as a quadratic hypersurface $\IP^5[2]$ in $\IP^5$. One can verify that this relation extends to a duality of the quantum K-theory at $t=0$, by establishing the equality of the level zero $I$-functions
\be
I_{\Gr(2,4)}(0)\simeq I_{\IP^5[2]}(0)=(1-q)\sum_{d\geq 0}Q^\td\frac{\prod_{r=1}^{2d}(1-q^{r-2\eps})}{\prod_{r=1}^d (1-q^{r-\eps})^6}\ .
\ee
The easiest way to show this equation is to check that the r.h.s.  satisfies the same difference equation from Tab.~\ref{tab:diffop} and to compare the first terms in the $Q$-expansion to fix the normalization. 

For the hypersurface in $\IP^5$, we choose the basis $\Phi_k=(1-e^{-H})^k$, $k=0,...,4$, where $H$ denotes the hyperplane class. By the classical duality it is related to the basis $\{\cx O_\mu\}$ for $K(\Gr(2,4))$ by $\Phi_k=(\cx O_1)^k$. Because of the symmetry exchanging $\cx O_2$ and $\cx O_{1,1}$ at $t=0$, one can reduce the six dimensional basis  $\{\cx O_\mu\}$ to the five-dimensional basis $\{(\cx O_1)^k\}$, as long as the direction $\cx O_{2}-\cx O_{1,1}$ stays undeformed. The structure constants at $t=0$ in the basis $\{(\cx O_1)^k\}$ are 
\be
C_{\cx O_1,\{\cx O_\mu\}} = 
\begin{small}\begin{pmatrix}
0&1&0&0&0&0\\
0&0&1&1&-1&0\\
0&0&0&0&1&0\\
0&0&0&0&1&0\\
Q&-Q&0&0&0&1\\
0&Q&0&0&0&0
\end{pmatrix}\end{small}
\ \to\
C_{\Phi_1,\{\Phi_k\}} =\begin{small}
 \begin{pmatrix}
0&1&0&0&0\\
0&0&1&0&0\\
-Q&Q&0&1&0\\
2Q&-3Q&0&0&1\\
0&2Q&0&0&0
\end{pmatrix} \  .\end{small}
\ee
One can check that the r.h.s. agrees with the structure constants at $t=0$ computed from $I_{\IP^5[2]}(0)$. 

One way to reconstruct the perturbed theory is to integrate the flatness equations \eqref{fleq}, as described in ref.~\cite{IMT}. This has the advantage of getting easily all order expressions in the general perturbation\footnote{The dependence on $T_0$ is fixed by the string equation \cite{Lee:2001mb},\cite{Giv15all}(p.VII).} $T=\sum_{k=1}^4T_k\Phi_k$ at fixed power of $Q$. Restricting to the direction $t=T_1$, which is the integrable deformation in the 2d limit, the perturbed structure constants at order $Q^1$ obtained in this way are
\be\label{Cphi1}
C_{\Phi_1}(t)|_{Q^1} =\begin{small}
Qe^t\left(
\begin{array}{ccccc}
 0 & 0 & 0 & 0 & 0 \\
 0 & 0 & 0 & 0 & 0 \\
 -1  &   (t+1) & -\frac{1}{2}   t (t+2) & \frac{1}{6}   t^2 (t+3) & -\frac{1}{24}   t^3 (t+4) \\
 2   & -  (2 t+3) &   t (t+3) & -\frac{1}{6}   t^2 (2 t+9) & \frac{1}{12}   t^3 (t+6) \\
 0 & 2   & -2   t &   t^2 & -\frac{1}{3}   t^3 \\
\end{array}
\right)\ . \end{small}
\ee
In the perturbed theory, the structure constants get contributions also  from higher order in $Q$. Some higher order terms are given in eq.~\eqref{Cp52}.

\bigskip
\noindent{\bf Acknowledgments:} 
We would like to thank 
Tudor Dimofte, 
Heeyeon Kim,
Nikita Nekrasov,
Yongbin Ruan,
Eric Sharpe,
Chris Woodward,
and
Ming Zhang 
for discussions and correspondences.
The work of P.M. is supported by the German Excellence Cluster Origins. U.N. is supported by the graduate school BCGS, and A.T. is supported by the DFG.\\[-1cm]

\bigskip

\bigskip

\appendix
\section{Appendix}
\subsection{Perturbed structure constants for $\IP^5(2) \simeq \Gr(2,4)$}
The perturbed structure constants \eqref{Cphi1} in the direction $t\Phi_1$ up to order $Q^3$ are:
\begin{small}
\bea \label{Cp52} &Q^0:&
\left(
\begin{array}{ccccc}
 0 & 1 & 0 & 0 & 0 \\
 0 & 0 & 1 & 0 & 0 \\
 0 & 0 & 0 & 1 & 0 \\
 0 & 0 & 0 & 0 & 1 \\
 0 & 0 & 0 & 0 & 0 \\
\end{array}
\right)
\nonumber \\
&Q^1:&\left(
\begin{array}{ccccc}
 0 & 0 & 0 & 0 & 0 \\
 0 & 0 & 0 & 0 & 0 \\
 \frac{1}{6} \left(-t^3-3 t^2-6 t-6\right) & \frac{2 t^3}{3}+\frac{3 t^2}{2}+2 t+1 & -\frac{1}{2} t \left(2
   t^2+3 t+2\right) & \frac{1}{6} t^2 (4 t+3) & -\frac{t^3}{6} \\
 \frac{t^3}{3}+t^2+2 t+2 & -\frac{3 t^3}{2}-\frac{7 t^2}{2}-5 t-3 & \frac{1}{2} t \left(5 t^2+8 t+6\right) &
   -\frac{1}{6} t^2 (11 t+9) & \frac{t^3}{2} \\
 0 & \frac{t^3}{3}+t^2+2 t+2 & -t \left(t^2+2 t+2\right) & t^2 (t+1) & -\frac{t^3}{3} \\
\end{array}
\right)+O(t^4)\, ,\nonumber \\
&Q^2:&\left(
\begin{array}{ccccc}
 0 & 0 & 0 & 0 & 0 \\
 \frac{t^4}{24} & -\frac{t^4}{12} & \frac{t^4}{24} & 0 & 0 \\
 -\frac{1}{24} t^2 \left(53 t^2+36 t+12\right) & \frac{1}{24} t^2 \left(113 t^2+56 t+12\right) & -\frac{2}{3}
   t^3 (4 t+1) & \frac{5 t^4}{12} & 0 \\
 \frac{1}{12} t^2 \left(71 t^2+52 t+18\right) & -\frac{1}{24} t^2 \left(309 t^2+164 t+36\right) & \frac{1}{2}
   t^3 (15 t+4) & -\frac{5 t^4}{4} & 0 \\
 -\frac{1}{3} t^2 \left(10 t^2+8 t+3\right) & \frac{1}{12} t^2 \left(91 t^2+52 t+12\right) & -\frac{2}{3} t^3
   (7 t+2) & \frac{5 t^4}{6} & 0 \\
\end{array}
\right)+O(t^5)\nonumber \\
&Q^3:&\left(
\begin{array}{ccccc}
 0 & 0 & 0 & 0 & 0 \\
 0 & 0 & 0 & 0 & 0 \\
 -\frac{5 t^4}{12} & \frac{5 t^4}{12} & 0 & 0 & 0 \\
 \frac{5 t^4}{4} & -\frac{5 t^4}{4} & 0 & 0 & 0 \\
 -\frac{5 t^4}{6} & \frac{5 t^4}{6} & 0 & 0 & 0 \\
\end{array}
\right)+O(t^5)
\eea
\end{small}
\subsection{Quantum K-theory for the dual pair $\Gr(2,5)\simeq \Gr(3,5)$ \label{sec:appG25}}
The Grassmannians $\Gr(2,5)$ and $\Gr(3,5)$ are related by the well-known classical geometric duality. In the following we compute the quantum K-theoretic product and test the duality at the quantum level.

The quantum product for the K-theoretic Schur polynomials $\sigma_\mu(x^K)$ in $\Gr(2,5)$ can be derived from the difference operators eq.~\eqref{Ican} along the lines of sect.~\ref{sec:WLA}. The K-theoretic Schur polynomials $\sigma_\mu(x^K)$ are related to the Grothedieck polynomials $\cx O_\mu(x^K)$ for $\Gr(2,5)$ by\\[-4mm] 
\be\label{baschangeGr25}
\hbox{
\vbox{\offinterlineskip
\halign{\strut~$#$~~\hfil&=~$#$\,,\hfil\hskip1cm&
~$#$~~\hfil&=~$#$\,,\hfil\hskip1cm&
~$#$~~\hfil&=~$#$\hfil\cr
\cx O_1  &\sigma_1 -\sigma_{1,1} &
\cx O_2  &\sigma_2 -\sigma_{2,1}&
\cx O_{1,1}  & \sigma_{1,1}\,,\cr
\cx O_{3} &\sigma_{3}-\sigma_{3,1}&
\cx O_{2,1}  & \sigma_{2,1} - \sigma_{2,2}&
\cx O_{3,1}  &\sigma_{3,1}-\sigma_{3,2}\,,\cr
\cx O_{2,2} & \sigma_{2,2}&
\cx O_{3,2} & \sigma_{3,2}-\sigma_{3,3}&
\cx O_{3,3} & \sigma_{3,3}\,.\cr
}}}
\ee
In the basis $\{\cx O_\mu\}$ with the given ordering, the inner product \eqref{ip} is
\begin{small}
\be\label{ipGr25}
\chi(\cx O_\mu,\cx O_\nu) = \begin{pmatrix}
	1 & 1 & 1 & 1 & 1 & 1 & 1 & 1 & 1 & 1 \\[-1.5mm]
	1 & 1 & 1 & 1 & 1 & 1 & 1 & 1 & 1 & 0 \\[-1.5mm]
	1 & 1 & 1 & 1 & 1 & 1 & 1 & 0 & 0 & 0 \\[-1.5mm]
	1 & 1 & 1 & 1 & 0 & 1 & 0 & 1 & 0 & 0 \\[-1.5mm]
	1 & 1 & 1 & 0 & 1 & 0 & 0 & 0 & 0 & 0 \\[-1.5mm]
	1 & 1 & 1 & 1 & 0 & 1 & 0 & 0 & 0 & 0 \\[-1.5mm]
	1 & 1 & 1 & 0 & 0 & 0 & 0 & 0 & 0 & 0 \\[-1.5mm]
	1 & 1 & 0 & 1 & 0 & 0 & 0 & 0 & 0 & 0 \\[-1.5mm]
	1 & 1 & 0 & 0 & 0 & 0 & 0 & 0 & 0 & 0 \\[-1.5mm]
	1 & 0 & 0 & 0 & 0 & 0 & 0 & 0 & 0 & 0 
\end{pmatrix},\qquad \det(\chi)=1\ .
\ee\end{small}
\hskip-6pt Written in the basis of the structure sheaves, the multiplication table becomes
\begin{small}
\bea\label{Tab25}
&&\hskip-0.9cm
\begin{array}{c|cccc}		
		*& \cx O_{1} & \cx O_{2} & \cx O_{1,1} & \cx O_{3}   \\
		\noalign{\hrule}
		\cx O_{1} & \cx O_{1,1}\!-\!\cx O_{2,1}\!+\!\cx O_{2} & -& -& -
		
		\\ \cx O_{2}  & \cx O_{2,1}\!-\!\cx O_{3,1}\!+\!\cx O_{3} & \cx O_{2,2}\!+\!\cx O_{3,1}\!-\!\cx O_{3,2} & -& -
		
		\\ \cx O_{1,1}  & \cx O_{2,1} & \cx O_{3,1} &  \cx O_{2,2} & -
		
		\\ \cx O_{3}  & \cx O_{3,1} & \cx O_{3,2} & Q & \cx O_{3,3} 
		
		\\ \cx O_{2, 1}  & \cx O_{2,2}\!+\!\cx O_{3,1}\!-\!\cx O_{3,2} & \cx O_{3,2}\!+\!Q \left(1\!-\!\cx O_{1}\right) & \cx O_{3,2} & Q \cx O_{1} 
		
		\\ \cx O_{3, 1} &\cx O_{3,2}\!+\!Q \left(1\!-\!\cx O_{1}\right) & \cx O_{3,3}\!+\!Q \left(\cx O_{1}\!-\!\cx O_{2}\right) & Q \cx O_{1} & Q \cx O_{2} 
		
		\\ \cx O_{2, 2}  &  \cx O_{3,2} & Q \cx O_{1} & \cx O_{3,3} &  Q \cx O_{1,1} 
		
		\\ \cx O_{3, 2}  & \cx O_{3,3}\!+\!Q \left(\cx O_{1}\!-\!\cx O_{2}\right) & Q \left(\cx O_{1,1}\!-\!\cx O_{2,1}\!+\!\cx O_{2}\right) & Q \cx O_{2} & Q \cx O_{2,1} 
		
		\\ \cx O_{3, 3} & Q \cx O_{2} & Q \cx O_{2,1} & Q \cx O_{3} & Q \cx O_{2,2} 
	\end{array}
\\\vspace{.5cm}
&&\hskip-0.9cm\begin{array}{c|ccccc}		
		*&  \cx O_{2, 1} & \cx O_{3, 1} & \cx O_{2, 2} & \cx O_{3, 2} & \cx O_{3, 3}  \\
		\noalign{\hrule}
		\cx O_{1} &  -& -& -& -& -
		
		\\ \cx O_{2}  &  -& -& -& -& -
		
		\\ \cx O_{1,1}  &  -& -& -& -& -
		
		\\ \cx O_{3}  &  -& -& -& -& -
		
		\\ \cx O_{2, 1}  & \cx O_{3,3}\!+\!Q \left(\cx O_{1}\!-\!\cx O_{2}\right) & -& -& -& -
		
		\\ \cx O_{3, 1} &Q \left(\cx O_{1,1}\!-\!\cx O_{2,1}\!+\!\cx O_{2}\right) & Q \left(\cx O_{2,1}\!-\!\cx O_{3,1}\!+\!\cx O_{3}\right) & -& -& -
		
		\\ \cx O_{2, 2}  &   Q \cx O_{2} & Q \cx O_{2,1} & Q \cx O_{3} & -& -
		
		\\ \cx O_{3, 2}  & Q \left(\cx O_{2,1}\!-\!\cx O_{3,1}\!+\!\cx O_{3}\right) & Q \left(\cx O_{2,2}\!+\!\cx O_{3,1}\!-\!\cx O_{3,2}\right) & Q \cx O_{3,1} & Q \cx O_{3,2}\!+\!Q^2 \left(1\!-\!\cx O_{1}\right) & -
		
		\\ \cx O_{3, 3} & Q \cx O_{3,1} & Q \cx O_{3,2} &  Q^2 & Q^2 \cx O_{1} &  Q^2 \cx O_{1,1}
	\end{array}\nonumber
\eea
\end{small}
\hskip-6pt The above multiplications agree with those derived from the quantum Pieri and Giambelli rules of ref.~\cite{BM08}. Moreover, the above  quantum product can be shown to be isomorphic to the  polynomial ring 
\bea
&&\hskip-1.7cm\mathbb{Z}[\cx O_{1} ,\cx O_{1,1} ,Q]/ (r_1 ,r_2)\, ,\nonumber\\[2mm]
r_1&=&\cx O_1^4+2 \cx O_1^3 \cx O_{1,1}+\cx O_1^2 \cx O_{1,1}^2-3
   \cx O_1^2 \cx O_{1,1}-2 \cx O_1\cx O_{1,1}^2+\cx O_{1,1}^2\, ,\\
r_2&=&\cx O_1^3
   \cx O_{1,1}+2 \cx O_1^2 \cx O_{1,1}^2+\cx O_1 \cx O_{1,1}^3-2 \cx O_1
   \cx O_{1,1}^2-\cx O_{1,1}^3- Q\, .\nonumber
\eea
The product determines the difference operator $\mx L$ annihilating the $I$-function as discussed in sect.~\ref{sec:DiffeqSec}. Writing $\mx L = \sum_{i\geq0}Q^i\mx L_i$ one finds the operator of degree ${5 \choose 2}=$10 in $\delta$ and degree 4 in $Q$:
\begin{small}
\bea\label{dop25}
\mx L_0 &=& -25 \delta ^7 \left(q^2+3 q+1\right)^3 (\delta +q-1)^3\, ,\nonumber\\
\mx L_1 &=&25 \delta ^{10} (q-1) q^4 (q+1) \left(q^2+q+1\right)-5 \delta ^6 q^4 \left(q^2+3 q+1\right)^3
   \left(63 q^2+10 q+2\right)
\nonumber\\&&
+5 \delta ^5 q^4 \left(q^2+3 q+1\right)^3 \left(78 q^2-16
   q-7\right)-25 \delta ^4 (q-1) q^4 (9 q+2) \left(q^2+3 q+1\right)^3
\nonumber\\&&+25 \delta ^3 (q-1)^2
   q^3 (2 q+1) \left(q^2+3 q+1\right)^3+5 \delta ^9 (q-1) q^4 \left(2 q^4+2 q^3-3 q^2-13
   q-13\right)\nonumber\\&&-5 \delta ^8 (q-1) q^4 \left(24 q^5+47 q^4+32 q^3+12 q^2-3 q-12\right)\nonumber\\&&+5 \delta
   ^7 q^4 \left(25 q^8+191 q^7+677 q^6+1087 q^5+815 q^4+295 q^3+39 q^2-7 q+3\right)\, ,\nonumber\\
\mx L_2 &=&\delta ^{10} q^9-5 \delta ^8 (q-1) q^9 (q+1)+25 q^8 \left(q^2+3 q+1\right)^2 \left(q^4+3
   q^2+1\right)\\&&-5 \delta ^7 (q-1) q^9 \left(25 q^4+50 q^3+51 q^2+28 q+3\right)+5 \delta ^6
   (q-1) q^9 \left(115 q^4+190 q^3+162 q^2+79 q+4\right)\nonumber\\&&-25 \delta  q^8 \left(q^2+3
   q+1\right) \left(5 q^6+18 q^5+19 q^4+45 q^3+21 q^2+12 q+5\right)\nonumber\\&&-\delta ^5 q^8 \left(25
   q^8+225 q^7+1699 q^6+875 q^5+205 q^4-5 q^3-125 q^2+201 q+25\right)\nonumber\\&&+5 \delta ^4 q^8
   \left(25 q^8+210 q^7+766 q^6+598 q^5+685 q^4+525 q^3+200 q^2+104 q+12\right)\nonumber\\&&-5 \delta ^3
   q^8 \left(50 q^8+390 q^7+1096 q^6+1193 q^5+1645 q^4+1190 q^3+475 q^2+179 q+32\right)\nonumber\\&&+25
   \delta ^2 q^8 \left(10 q^8+72 q^7+180 q^6+240 q^5+345 q^4+240 q^3+108 q^2+45 q+10\right)\, ,\nonumber\\
\mx L_3 &=&-5 \delta ^7 q^{15}+5 \delta ^6 (5 q-2) q^{14}+\delta ^5 \left(25 q^5+25 q^4-70 q^2+34
   q-25\right) q^{13}\nonumber\\&&-5 \delta ^4 (q-1) \left(25 q^4+48 q^3+48 q^2+21 q+25\right) q^{13}+5
   \delta ^3 (q-1) \left(50 q^4+92 q^3+93 q^2+61 q+50\right) q^{13}\nonumber\\&&-5 \delta ^2 (q-1)
   \left(50 q^4+88 q^3+91 q^2+70 q+50\right) q^{13}+5 \delta  (q-1) \left(25 q^4+42 q^3+45
   q^2+38 q+25\right) q^{13}\nonumber\\&&-5 (q-1) \left(5 q^4+8 q^3+9 q^2+8 q+5\right) q^{13}\, ,\nonumber\\
\mx L_4 &=&\delta ^5 q^{19}-5 \delta ^4 q^{19}+10 \delta ^3 q^{19}-10 \delta ^2 q^{19}+5 \delta 
   q^{19}-q^{19}\ .\nonumber
\eea
\end{small}

The computation for the dual case $\Gr(3,5)$ is similar. The relation of structure sheaves $\cx O_\mu(x^K)$ and the Schur basis $\sigma_{\mu}(x^K)$ is now
\be
\hbox{
\vbox{\offinterlineskip
\halign{\strut~$#$~~\hfil&=~$#$\,,\hfil\hskip1.3cm&
~$#$~~\hfil&=~$#$\hfil\cr
\cx O_1  & \sigma_1 -\sigma_{1,1}+\sigma_{1,1,1} &
\cx O_2  & \sigma_2 -\sigma_{2,1}+\sigma_{2,1,1}\,  ,\cr
\cx O_{1,1}  & \sigma_{1,1}-2\sigma_{1,1,1}&
\cx O_{2,1} & \sigma_{2,1}-\sigma_{2,2}-2\sigma_{2,1,1}+2\sigma_{2,2,1}-\sigma_{2,2,2}\, ,\cr
\cx O_{1,1,1}  & \sigma_{1,1,1}&
\cx O_{2,2}  & \sigma_{2,2}-2\sigma_{2,2,1}+\sigma_{2,2,2}\,  ,\cr
\cx O_{2,1,1} & \sigma_{2,1,1}-\sigma_{2,2,1}+\sigma_{2,2,2} &
\cx O_{2,2,1} & \sigma_{2,2,1}-2\sigma_{2,2,2} \,  ,\quad 
\cx O_{2,2,2} = \sigma_{2,2,2} \, .\cr
}}}
\ee
The inner product on the basis of structure sheaves is
\begin{small}
\be\label{ipGr35}
\chi(\cx O_\mu,\cx O_\nu) =
\begin{pmatrix}	
	1 & 1 & 1 & 1 & 1 & 1 & 1 & 1 & 1 & 1 \\[-1.5mm]
	1 & 1 & 1 & 1 & 1 & 1 & 1 & 1 & 1 & 0 \\[-1.5mm]
	1 & 1 & 1 & 1 & 1 & 0 & 1 & 0 & 0 & 0 \\[-1.5mm]
	1 & 1 & 1 & 1 & 1 & 1 & 0 & 1 & 0 & 0 \\[-1.5mm]
	1 & 1 & 1 & 1 & 1 & 0 & 0 & 0 & 0 & 0 \\[-1.5mm]
	1 & 1 & 0 & 1 & 0 & 1 & 0 & 0 & 0 & 0 \\[-1.5mm]
	1 & 1 & 1 & 0 & 0 & 0 & 0 & 0 & 0 & 0 \\[-1.5mm]
	1 & 1 & 0 & 1 & 0 & 0 & 0 & 0 & 0 & 0 \\[-1.5mm]
	1 & 1 & 0 & 0 & 0 & 0 & 0 & 0 & 0 & 0 \\[-1.5mm]
	1 & 0 & 0 & 0 & 0 & 0 & 0 & 0 & 0 & 0 
\end{pmatrix},\qquad \det(\chi)=1\ .
\ee\end{small}
\hskip-6pt Taking acount a sign change $Q\to -Q$, which is due to our sign convention explained in footnote \ref{fnsign}, the quantum multiplication table for $\Gr(3,5)$ computed by polynomial reduction is equal to the one obtained from \eqref{Tab25} by transposing Young tableaus labelling the Grothendieck polynomials, i.e., $\cx O_\mu \to \cx O_{\mu^T}$. Moreover, the difference operator for $\Gr(3,5)$ coincides with that in eq.~\eqref{dop25}, which implies that the expansion coefficients \eqref{Iexpansion} of the $I$-functions agree up to linear combination.

From the point of the underlying 3d field theory, the agreement of the level zero quantum K-theory for $\Gr(2,5)$ and $\Gr(3,5)$ is expected in view of the Seiberg type dualities studied in ref.~\cite{BCC}. The agreement of the $I$-functions is the statement, that the $D^2\times S^1$ partition functions for these 3d theories agree upon appropriate identification of the vacua on both sides. From the algebraic point, the isomorphism between the Wilson line algebras is less obvious, and the vacuum equations $e^{d \widetilde W} =1$ of the Bethe/gauge correspondence of ref.~\cite{NekSha2} are different in the two theories. Non-trivial equivalences of algebras of this type have been discussed ref.~\cite{KW13}.


\begin{thebibliography}{10}

\bibitem{WittenVer}
E.~Witten, \emph{The {V}erlinde algebra and the cohomology of the
  {G}rassmannian},  in \emph{Geometry, topology, \& physics}, Conf. Proc.
  Lecture Notes Geom. Topology, IV, pp.~357--422.
\newblock Int. Press, Cambridge, MA, 1995.
\newblock \href{https://arxiv.org/abs/hep-th/9312104}{{\ttfamily
  hep-th/9312104}}.

\bibitem{VL88}
E.~P. Verlinde, \emph{{Fusion Rules and Modular Transformations in 2D Conformal
  Field Theory}},
  \href{https://doi.org/10.1016/0550-3213(88)90603-7}{\emph{Nucl. Phys.}
  {\bfseries B300} (1988) 360--376}.

\bibitem{Vafa:1990mu}
C.~Vafa, \emph{{Topological Landau-Ginzburg models}},
  \href{https://doi.org/10.1142/S0217732391000324}{\emph{Mod. Phys. Lett.}
  {\bfseries A6} (1991) 337--346}.

\bibitem{KI91}
K.~A. Intriligator, \emph{{Fusion residues}},
  \href{https://doi.org/10.1142/S0217732391004097}{\emph{Mod. Phys. Lett.}
  {\bfseries A6} (1991) 3543--3556},
  [\href{https://arxiv.org/abs/hep-th/9108005}{{\ttfamily hep-th/9108005}}].

\bibitem{CV92}
S.~Cecotti and C.~Vafa, \emph{{On classification of N=2 supersymmetric
  theories}}, \href{https://doi.org/10.1007/BF02096804}{\emph{Commun. Math.
  Phys.} {\bfseries 158} (1993) 569--644},
  [\href{https://arxiv.org/abs/hep-th/9211097}{{\ttfamily hep-th/9211097}}].

\bibitem{Nekphd}
N.~Nekrasov, \emph{{Four Dimensional Holomorphic Theories}}, Ph.D. thesis,
  Princeton University, 1996,
  {\href{http://scgp.stonybrook.edu/people/faculty/bios/nikita-nekrasov}{http://scgp.stonybrook.edu/people/faculty/bios/nikita-nekrasov}}.

\bibitem{KW13}
A.~Kapustin and B.~Willett, \emph{{Wilson loops in supersymmetric
  Chern-Simons-matter theories and duality}},
  \href{https://arxiv.org/abs/1302.2164}{{\ttfamily 1302.2164}}.

\bibitem{NN17}
N.~Nekrasov, \emph{{BPS/CFT correspondence IV: sigma models and defects in
  gauge theory}}, \href{https://doi.org/10.1007/s11005-018-1115-7}{\emph{Lett.
  Math. Phys.} {\bfseries 109} (2019) 579--622},
  [\href{https://arxiv.org/abs/1711.11011}{{\ttfamily 1711.11011}}].

\bibitem{NekSha}
N.~A. Nekrasov and S.~L. Shatashvili, \emph{{Quantization of Integrable Systems
  and Four Dimensional Gauge Theories}},  in \emph{{Proceedings, 16th
  International Congress on Mathematical Physics (ICMP09): Prague, Czech
  Republic, August 3-8, 2009}}, pp.~265--289, 2009,
  \href{https://arxiv.org/abs/0908.4052}{{\ttfamily 0908.4052}},
  \href{https://doi.org/10.1142/9789814304634_0015}{DOI}.

\bibitem{NekSha2}
N.~A. {Nekrasov} and S.~L. {Shatashvili}, \emph{{Supersymmetric Vacua and Bethe
  Ansatz}},
  \href{https://doi.org/10.1016/j.nuclphysbps.2009.07.047}{\emph{Nuclear
  Physics B Proceedings Supplements} {\bfseries 192} (July, 2009) 91--112},
  [\href{https://arxiv.org/abs/0901.4744}{{\ttfamily 0901.4744}}].

\bibitem{Giv15all}
A.~Givental, \emph{{Permutation-equivariant quantum K-theory I--XI}},  {
  [I~\href{https://arxiv.org/abs/1508.02690}{1508.02690}],
  [II~\href{https://arxiv.org/abs/1508.04374}{1508.04374}],
  [III~\href{https://arxiv.org/abs/1508.06697}{1508.06697}],
  [IV~\href{https://arxiv.org/abs/1509.00830}{1509.00830}],
  [V~\href{https://arxiv.org/abs/1509.03903}{1509.03903}],
  [VI~\href{https://arxiv.org/abs/1509.07852}{1509.07852}],
  [VII~\href{https://arxiv.org/abs/1510.03076}{1510.03076}],
  [VIII~\href{https://arxiv.org/abs/1510.06116}{1510.06116}],
  [IX~\href{https://arxiv.org/abs/1709.03180}{1709.03180}],
  [X~\href{https://arxiv.org/abs/1710.02376}{1710.02376}],
  [XI~\href{https://arxiv.org/abs/1711.04201}{1711.04201}]}, 2015-2017,
  \href{https://math.berkeley.edu/~giventh/perm/perm.html}{https://math.berkeley.edu/~giventh/perm/perm.html}.

\bibitem{JM}
H.~Jockers and P.~Mayr, \emph{{A 3d Gauge Theory/Quantum K-Theory
  Correspondence}},  \href{https://arxiv.org/abs/1808.02040}{{\ttfamily
  1808.02040}}.

\bibitem{RZ18}
Y.~{Ruan} and M.~{Zhang}, \emph{{The level structure in quantum K-theory and
  mock theta functions}},  \href{https://arxiv.org/abs/1804.06552}{{\ttfamily
  1804.06552}}.

\bibitem{RZ18b}
Y.~Ruan and M.~Zhang, \emph{{Verlinde/Grassmannian Correspondence and Rank 2
  $\delta$-wall-crossing}},  \href{https://arxiv.org/abs/1811.01377}{{\ttfamily
  1811.01377}}.

\bibitem{BM08}
A.~S. Buch and L.~C. Mihalcea, \emph{Quantum {$K$}-theory of {G}rassmannians},
  \href{https://doi.org/10.1215/00127094-2010-218}{\emph{Duke Math. J.}
  {\bfseries 156} (2011) 501--538},
  [\href{https://arxiv.org/abs/0810.0981}{{\ttfamily 0810.0981}}].

\bibitem{NekWit}
N.~Nekrasov and E.~Witten, \emph{{The Omega Deformation, Branes, Integrability,
  and Liouville Theory}},
  \href{https://doi.org/10.1007/JHEP09(2010)092}{\emph{JHEP} {\bfseries 09}
  (2010) 092}, [\href{https://arxiv.org/abs/1002.0888}{{\ttfamily 1002.0888}}].

\bibitem{GivEGW}
A.~Givental, \emph{Equivariant {G}romov-{W}itten invariants},
  \href{https://doi.org/10.1155/S1073792896000414}{\emph{Internat. Math. Res.
  Notices} (1996) 613--663}, [\href{https://arxiv.org/abs/9603021}{{\ttfamily
  9603021}}].

\bibitem{Dimofte:2011py}
T.~Dimofte, D.~Gaiotto and S.~Gukov, \emph{{3-Manifolds and 3d Indices}},
  \href{https://doi.org/10.4310/ATMP.2013.v17.n5.a3}{\emph{Adv. Theor. Math.
  Phys.} {\bfseries 17} (2013) 975--1076},
  [\href{https://arxiv.org/abs/1112.5179}{{\ttfamily 1112.5179}}].

\bibitem{BDP}
C.~Beem, T.~Dimofte and S.~Pasquetti, \emph{{Holomorphic Blocks in Three
  Dimensions}}, \href{https://doi.org/10.1007/JHEP12(2014)177}{\emph{JHEP}
  {\bfseries 12} (2014) 177},
  [\href{https://arxiv.org/abs/1211.1986}{{\ttfamily 1211.1986}}].

\bibitem{GivTon}
A.~Givental and V.~Tonita, \emph{The {H}irzebruch-{R}iemann-{R}och theorem in
  true genus-0 quantum {K}-theory},  in \emph{Symplectic, {P}oisson, and
  noncommutative geometry}, vol.~62 of \emph{Math. Sci. Res. Inst. Publ.},
  pp.~43--91.
\newblock Cambridge Univ. Press, New York, 2014.
\newblock \href{https://arxiv.org/abs/1106.3136}{{\ttfamily 1106.3136}}.

\bibitem{GivLee}
A.~Givental and Y.-P. Lee, \emph{Quantum {$K$}-theory on flag manifolds,
  finite-difference {T}oda lattices and quantum groups},
  \href{https://doi.org/10.1007/s00222-002-0250-y}{\emph{Invent. Math.}
  {\bfseries 151} (2003) 193--219},
  [\href{https://arxiv.org/abs/math/0108105}{{\ttfamily math/0108105}}].

\bibitem{Gadde:2013wq}
A.~Gadde, S.~Gukov and P.~Putrov, \emph{{Walls, Lines, and Spectral Dualities
  in 3d Gauge Theories}},
  \href{https://doi.org/10.1007/JHEP05(2014)047}{\emph{JHEP} {\bfseries 05}
  (2014) 047}, [\href{https://arxiv.org/abs/1302.0015}{{\ttfamily 1302.0015}}].

\bibitem{YS}
Y.~{Yoshida} and K.~{Sugiyama}, \emph{{Localization of 3d $\mathcal{N}=2$
  Supersymmetric Theories on $S^1 \times D^2$}},
  \href{https://arxiv.org/abs/1409.6713}{{\ttfamily 1409.6713}}.

\bibitem{DGP}
T.~Dimofte, D.~Gaiotto and N.~M. Paquette, \emph{{Dual boundary conditions in
  3d SCFTs}}, \href{https://doi.org/10.1007/JHEP05(2018)060}{\emph{JHEP}
  {\bfseries 05} (2018) 060},
  [\href{https://arxiv.org/abs/1712.07654}{{\ttfamily 1712.07654}}].

\bibitem{toappear}
H.~Jockers, P.~Mayr, U.~Ninad and A.~Tabler{\emph{{,\ to appear}} }.

\bibitem{CV13}
S.~Cecotti, D.~Gaiotto and C.~Vafa, \emph{{$tt^*$ geometry in 3 and 4
  dimensions}}, \href{https://doi.org/10.1007/JHEP05(2014)055}{\emph{JHEP}
  {\bfseries 05} (2014) 055},
  [\href{https://arxiv.org/abs/1312.1008}{{\ttfamily 1312.1008}}].

\bibitem{MR686357}
A.~Lascoux and M.-P. Sch\"{u}tzenberger, \emph{Structure de {H}opf de l'anneau
  de cohomologie et de l'anneau de {G}rothendieck d'une vari\'{e}t\'{e} de
  drapeaux}, {\emph{C. R. Acad. Sci. Paris S\'{e}r. I Math.} {\bfseries 295}
  (1982) 629--633}.

\bibitem{Wen}
Y.~{Wen}, \emph{{K-Theoretic $I$-function of $V//_{\theta} \mathbf{G}$ and
  Application}},  \href{https://arxiv.org/abs/1906.00775}{{\ttfamily
  1906.00775}}.

\bibitem{Tai}
K.~{Taipale}, \emph{{K-theoretic J-functions of type A flag varieties}},
  \href{https://arxiv.org/abs/1110.3117}{{\ttfamily 1110.3117}}.

\bibitem{Woodward:2018rab}
C.~Woodward and G.~Xu, \emph{{An open quantum Kirwan map}},
  \href{https://arxiv.org/abs/1806.06717}{{\ttfamily 1806.06717}}.

\bibitem{HV}
K.~Hori and C.~Vafa, \emph{{Mirror symmetry}},
  \href{https://arxiv.org/abs/hep-th/0002222}{{\ttfamily hep-th/0002222}}.

\bibitem{Martin}
S.~Martin, \emph{Symplectic quotients by a nonabelian group and by its maximal
  torus},  \href{https://arxiv.org/abs/math/0001002}{{\ttfamily math/0001002}}.

\bibitem{GivER}
A.~Givental, \emph{Explicit reconstruction in quantum cohomology and
  {K}-theory}, \href{https://doi.org/10.5802/afst.1500}{\emph{Ann. Fac. Sci.
  Toulouse Math. (6)} {\bfseries 25} (2016) 419--432},
  [\href{https://arxiv.org/abs/1506.06431}{{\ttfamily 1506.06431}}].

\bibitem{KW09}
A.~Kapustin, B.~Willett and I.~Yaakov, \emph{{Exact Results for Wilson Loops in
  Superconformal Chern-Simons Theories with Matter}},
  \href{https://doi.org/10.1007/JHEP03(2010)089}{\emph{JHEP} {\bfseries 03}
  (2010) 089}, [\href{https://arxiv.org/abs/0909.4559}{{\ttfamily 0909.4559}}].

\bibitem{BZ}
F.~Benini and A.~Zaffaroni, \emph{{Supersymmetric partition functions on
  Riemann surfaces}}, {\emph{Proc. Symp. Pure Math.} {\bfseries 96} (2017)
  13--46}, [\href{https://arxiv.org/abs/1605.06120}{{\ttfamily 1605.06120}}].

\bibitem{ClK}
C.~Closset and H.~Kim, \emph{{Comments on twisted indices in 3d supersymmetric
  gauge theories}}, \href{https://doi.org/10.1007/JHEP08(2016)059}{\emph{JHEP}
  {\bfseries 08} (2016) 059},
  [\href{https://arxiv.org/abs/1605.06531}{{\ttfamily 1605.06531}}].

\bibitem{IS}
K.~Intriligator and N.~Seiberg, \emph{{Aspects of 3d N=2 Chern-Simons-Matter
  Theories}}, \href{https://doi.org/10.1007/JHEP07(2013)079}{\emph{JHEP}
  {\bfseries 07} (2013) 079},
  [\href{https://arxiv.org/abs/1305.1633}{{\ttfamily 1305.1633}}].

\bibitem{GivWDVV}
A.~Givental, \emph{On the {WDVV} equation in quantum {$K$}-theory},
  \href{https://doi.org/10.1307/mmj/1030132720}{\emph{Michigan Math. J.}
  {\bfseries 48} (2000) 295--304},
  [\href{https://arxiv.org/abs/math/0003158}{{\ttfamily math/0003158}}].

\bibitem{Lee:2001mb}
Y.-P. Lee, \emph{Quantum {$K$}-theory. {I}. {F}oundations},
  \href{https://doi.org/10.1215/S0012-7094-04-12131-1}{\emph{Duke Math. J.}
  {\bfseries 121} (2004) 389--424},
  [\href{https://arxiv.org/abs/math/0105014}{{\ttfamily math/0105014}}].

\bibitem{IMT}
H.~Iritani, T.~Milanov and V.~Tonita, \emph{Reconstruction and convergence in
  quantum {$K$}-theory via difference equations},
  \href{https://doi.org/10.1093/imrn/rnu026}{\emph{Int. Math. Res. Not. IMRN}
  (2015) 2887--2937}, [\href{https://arxiv.org/abs/1309.3750}{{\ttfamily
  1309.3750}}].

\bibitem{BCC}
F.~Benini, C.~Closset and S.~Cremonesi, \emph{{Comments on 3d Seiberg-like
  dualities}}, \href{https://doi.org/10.1007/JHEP10(2011)075}{\emph{JHEP}
  {\bfseries 10} (2011) 075},
  [\href{https://arxiv.org/abs/1108.5373}{{\ttfamily 1108.5373}}].

\bibitem{Batetal}
V.~V. Batyrev, I.~Ciocan-Fontanine, B.~Kim and D.~van Straten, \emph{{Conifold
  transitions and mirror symmetry for Calabi-Yau complete intersections in
  Grassmannians}},
  \href{https://doi.org/10.1016/S0550-3213(98)00020-0}{\emph{Nucl. Phys.}
  {\bfseries B514} (1998) 640--666},
  [\href{https://arxiv.org/abs/alg-geom/9710022}{{\ttfamily
  alg-geom/9710022}}].

\end{thebibliography}
\providecommand{\href}[2]{#2}\begingroup\raggedright\endgroup
\end{document}